# Automated Repair of Declarative Software Specifications in the Era of Large Language Models

Md Rashedul Hasan, Jiawei Li, Iftekhar Ahmed, *Member, IEEE* and Hamid Bagheri, *Member, IEEE*

*Abstract*—The growing adoption of declarative software specification languages, coupled with their inherent difficulty in debugging, has underscored the need for effective and automated repair techniques applicable to such languages. Researchers have recently explored various methods to automatically repair declarative software specifications, such as template-based repair, feedback-driven iterative repair, and bounded exhaustive approaches. The latest developments in large language models provide new opportunities for the automatic repair of declarative specifications. In this study, we assess the effectiveness of utilizing OpenAI's ChatGPT to repair software specifications written in the Alloy declarative language. Unlike imperative languages, specifications in Alloy are not executed but rather translated into logical formulas and evaluated using backend constraint solvers to identify specification instances and counterexamples to assertions. Our evaluation focuses on ChatGPT's ability to improve the correctness and completeness of Alloy declarative specifications through automatic repairs. We analyze the results produced by ChatGPT and compare them with those of leading automatic Alloy repair methods. Our study revealed that while ChatGPT falls short in comparison to existing techniques, it was able to successfully repair bugs that no other technique could address. Our analysis also identified errors in ChatGPT's generated repairs, including improper operator usage, type errors, higherorder logic misuse, and relational arity mismatches. Additionally, we observed instances of hallucinations in ChatGPT-generated repairs and inconsistency in its results. Our study provides valuable insights for software practitioners, researchers, and tool builders considering ChatGPT for declarative specification repairs.

## I. INTRODUCTION

THE versatile nature of declarative specification languages has enabled their application in various software engineering problems. One such language is the Alloy specification language [36], which employs relational algebra and first-order logic to solve a broad range of tasks in software engineering. These tasks include software design [12]–[14], [37], [43], [83], software verification [10], [28], [58], [63], [67], [74], [85], network security [56], [64], [75], security analysis of emerging platforms, such as IoT and Android platforms [6], [8], [9], [11], [15], [77], test case generation [41], [59], [62], [70], [78], and more. Furthermore, the Alloy specification language is seamlessly integrated with the Alloy Analyzer, which enables users to automatically verify if the provided specifications meet certain desirable properties. This integration simplifies the analysis and verification of Alloy specifications and enables a seamless user experience within the Alloy environment. Although the Alloy Analyzer is a powerful tool, Alloy users, much like Java or C developers, may still inadvertently introduce subtle bugs while writing their specifications. Even though the Alloy Analyzer can assist in automatically verifying properties and producing counterexamples that highlight assertion violations, the actual process of debugging, evaluating, and resolving issues can still be difficult and timeconsuming, particularly for specifications that capture intricate behaviors and large systems [50], [57].

In recent years, researchers have explored using various methods to automatically repair declarative software specifications, such as template-based repair [93], [95], feedback-driven iterative repair [19], and bounded exhaustive approaches [18], [34]. ARepair [88] is considered the first approach for automated program repair in Alloy. It operates by fixing bugs identified by failing test cases, and thus assumes that tests are available to verify correctness. However, tests are not commonly used in Alloy specifications, where users rely on assertions to describe expectations and use the Alloy Analyzer to validate them. Moreover, ARepair can be susceptible to the *overfitting* issue when test cases are insufficient, resulting in generated repairs that pass the tests but are incorrect in general [19], similar to the problems encountered in traditional imperative-based automatic program repair (APR) approaches. BeAFix [34] technique utilizes assertions as correctness oracles and mitigates the overfitting problem. Nevertheless, the BeAFix approach's drawback is that it can be time-consuming as it exhaustively enumerates the possible candidate repairs up to a certain threshold and verifies if any of them successfully satisfy the failing assertions. A prior research [95] has shown that BeAFix is limited in generating complex repairs for a range of Alloy specifications as the number of repair candidates to be tried can be prohibitively large, indicating ample opportunity for improvement.

The advent of Large Language Models (LLMs) trained on massive amounts of natural language documents, code, and specifications has revolutionized the field of natural language processing and the applications thereof. By employing advanced architectures, such as transformers [20], [86], and leveraging the power of large datasets, these pre-trained models have demonstrated remarkable accuracy in predicting not only natural language but also code and specifications. Notably, LLMs have proven adept at tasks such as code completion and generating code from natural language descriptions [92]. The potential of LLMs has been demonstrated by Codex [21], one of the largest

M. Hasan and H. Bagheri are with the School of Computing, University of Nebraska-Lincoln, Lincoln, NE, USA. E-mail: {mhasan6,hbagheri}@cse.unl.edu

J. Li, I. Ahmed are with the Donald Bren School of Information and Computer Science, University of California, Irvine, USA. E-mail: {jiawl28,iftekha}@uci.edu

models which is now incorporated in Github Copilot[1], a practical tool designed to aid developers by producing code based on user input. ChatGPT[2] is another language model that has garnered significant attention since its public release in November 2022. Its intuitive interface allows users to interact with the model naturally, making it a valuable asset for software engineering tasks.

The well-established potential of ChatGPT in program generation and automation, combined with the widely-known challenges involved in repairing formal specifications, which is a time-consuming and laborious process for software developers, inspired us to conduct a systematic empirical study. The primary objective of our study is to explore the possibilities of using ChatGPT for repairing faulty Alloy specifications, and to identify any obstacles that must be addressed to make it a viable solution.

In this paper, we present and interpret the results of our comprehensive evaluation of ChatGPT's effectiveness in repairing faulty Alloy specifications, comparing it to the state-of-the-art Alloy APR techniques recently proposed in the literature. Our evaluation includes ARepair [88], BeAFIX [34], ATR [95], and ICEBAR [19] as baselines for comparison. We evaluate all the techniques on two sets of benchmarks, developed by other research groups, consisting of 1,974 defective specifications from ARepair [88] and Alloy4Fun [55]. We provide a categorization of the errors we discovered and discuss the implications of these findings for formal specification repair. Our research aims to address the following questions:

- **RQ1:** How does the performance of the LLM-based repair, realized atop ChatGPT, compare to that of the stateof-the-art techniques for repairing Alloy specifications?
- **RQ2:** How consistent are the repairs generated by the LLM-based repair, which is built on top of ChatGPT, when using the same prompt and faulty specifications?
- **RQ3:** What are the primary categories of mistakes present in ChatGPT's generated Alloy repairs?
- **RQ4:** What impact do prompts have on the Alloy repairs produced by ChatGPT?

To summarize, this paper makes the following contributions:

- We conduct the first empirical study, to the best of our knowledge, that explores the effectiveness of LLM-based Alloy repair, realized atop ChatGPT, in comparison to that of the state-of-the-art techniques for repairing Alloy specifications. This study becomes increasingly crucial with the prevalence of powerful AI tools.
- We offer a comprehensive analysis and interpretation of the results obtained, accompanied by a categorization of errors that arise from LLM-based repair compared to other techniques for Alloy APR.
- We provide a complete replication package to support the verifiability and replication of our study [1].

## II. BACKGROUND AND MOTIVATION

### A. Alloy Overview

Alloy is a formal modeling language with a comprehensible syntax inspired by object-oriented notations and a semantics based on first-order relational logic [3]. An Alloy specification consists of three key components: (1) data types, (2) formulas that define constraints over data types, and (3) commands to execute the analyzer. The language uses signatures (sig) to define basic data types, which can be structured using fields to capture the relationships between signatures.

Facts (fact) are formulas that take no arguments, and introduce constraints that always hold in every instance of the specification, thus restricting the model space of the specification. The formulas can be further structured using predicates (pred) and functions (fun), which are named, parameterized Alloy formulas that can be invoked. Assertions (assert) are named formulas that capture expected properties of the specification, i.e., properties that the user wants to verify. The check and run commands are used to invoke the automated analyzer for checking assertions (assert), which are Alloy formulas used to check properties, and running predicates (pred), respectively.

Formulas in Alloy specifications are expressed in relational logic, which is a first-order logic that has been extended with relational operators such as transpose, union, difference, and intersection. In addition to the usual universal and existential quantifiers (all and some, respectively), Alloy supports one and lone quantifiers, which respectively stand for "exists exactly one" and "exists at most one". Furthermore, Alloy features important relational operators such as relational join, which generalizes composition to n-ary relations and allows for the expression of navigations as in object orientation. Another significant operator is the transitive closure, which can be applied only to binary relations and extends the expressiveness of Alloy beyond that of first-order logic.

The Alloy Analyzer automates the analysis of Alloy specifications by translating them into propositional formulas and using off-the-shelf SAT solvers to find either counterexamples that violate asserted properties (with check command) or model instances that satisfy predicates (with run command). In fact, the analyzer performs two types of automated analysis: *simulation* and *model checking*.

Simulation demonstrates the consistency of model specifications by generating a satisfying model instance, while model checking involves finding a counterexample–a model instance that violates a particular assertion. When provided with a specification $M$ and an assertion $p$, the Alloy Analyzer searches for counterexamples by solving a propositional formula that represents the constraints of $M$ and the negation of $p$ ($M \land \neg p$). On the other hand, if $p$ is a predicate, the Alloy Analyzer will search for model instances that satisfy *p* by solving the formula $M \wedge p$. Alloy's analysis is bounded-exhaustive, meaning that if a specification *M*, an assertion (or a predicate) *p*, and a bound *t* (referred to as the *scope* in Alloy's context) are given, then the Alloy Analyzer will only find a counterexample (or a satisfying instance) for *p* of size that is bounded by *t*, if such a *t*-bounded counterexample (or instance) exists.

---

[1] https://github.com/features/copilot
[2] https://chat.openai.com/chat

```
1   sig Room {}
2   one sig SecureLab extends Room {}
3
4   abstract sig Person { owns: set Key }
5   sig Employee extends Person {}
6   sig Researcher extends Person {}
7
8   sig Key { opened_by: one Room }
9
10  fact {
11    some Employee && some Researcher
12    some e: Employee | #e.owns > 1
13  }
14
15  fact {
16    all r: Room | some opened_by.r
17    all e: Employee| some k: Key| k in e.owns
18    and SecureLab != k.opened_by  //bug
19    //and e !in owns.opened_by.SecureLab
20    //missing constraint
21  }
22
23  pred CanEnter(p: Person, r:Room) {
24      r in p.owns.opened_by
25  }
26
27  assert is_secured {
28    all p: Person | CanEnter[p, SecureLab]
29      implies p in Researcher
30  }
31  check is_secured
```

Fig. 1: A faulty Alloy specification.

### B. Illustrative Example

The Alloy specification presented in Fig. 1 addresses a room access control problem. The specification begins by defining several classes or types, called *sig*'s. The Room type has only one subclass, SecureLab. The Person type has two non-overlapping subclasses, Employee and Researcher, both sharing a common field, owns, which represents a set of keys owned by a person. The Key type has a field, opened_by, which represents exactly one room that can be opened by the key. Additionally, a couple of *facts* are defined in the specification to capture additional constraints. The first fact specifies that there must be at least one employee and one researcher, and at least one employee who owns more than one key.

The second *fact* in the Alloy specification for a room access control problem (Fig. 1, lines 15–21), specifies that every room can be opened by some key and that each employee owns at least one key, which cannot open SecureLab. In addition to these *facts*, there is a *predicate* called CanEnter, which is a boolean function that checks whether a person can enter a room based on whether they own a key to that room. Finally, to ensure the secure access of SecureLab, the *assert* is_secured is used to verify that only researchers can enter the secure lab.

To verify the assertion, the Alloy Analyzer is executed on this specification using the check command, as shown in line 31. The Analyzer generates a counterexample that violates the assertion by describing a scenario where an employee owns a key that can open SecureLab. This demonstrates that the specification contains a bug that violates the asserted property. The objective is to evaluate the effectiveness of a repair method using LLMs, such as ChatGPT, compared to the state-of-the-art Alloy repair for fixing such bugs.

## III. RELATED WORK

### A. Automated Program Repair

Automatic program repair has become an increasingly popular research topic due to the growing demand for reliable software. There are several different approaches to automatic program repair. Constraint-based repair techniques, such as SemFix [65], FoRenSiC [17], Angelix [60], AFix [38], StarFix [94], Jobstmann et al. [39], and Gopinath et al. [30], use constraints to produce repairs that are correct by construction, ensuring that they adhere to a given specification or pass a test suite. In contrast, generate-and-validate repair techniques, such as Pachika [23], PAR [42], Prophet [54], Debroy and Wong [24], and GenProg [47], find several repair candidates using stochastic search or invariant inferences, and then verify them against the given specifications. Learning-based repair techniques, such as iFixR [45], Fixminer [44], DLFix [51], and DeepDelta [61], learn fixes from repair samples.

Some research efforts in program synthesis and repair, such as [7], [46], [60], [65], [66], [81], [82], integrate existing tools like test-input generation or symbolic execution to synthesize desired programs. These integrations are common in modern synthesis techniques, including the multi-disciplinary ExCAPE project [4] and the SyGuS competition [5]. As a result, many practical and useful tools have been developed, such as Sketch [80], which generates low-level bit-stream programs, Autograder [79], which provides feedback on programming homework, and FlashFill [31], [32], which constructs Excel macros.

**Alloy Repair.** Compared to automatic program repair and synthesis approaches for imperative programs, there has been relatively less research dedicated to APR techniques for declarative specifications, such as Alloy.

ARepair [88] was the first APR approach for Alloy, which generates fixes for Alloy specifications violating test cases. The ARepair approach utilizes three techniques: AUnit [84] tests, which are unit tests expressed as Alloy predicates; AlloyFL [89] fault localizer, which uses unit tests to identify potentially problematic statements in the program; and mutation-based repairs (e.g., [47]), which randomly mutate Alloy expressions in an attempt to fix bugs. However, prior studies [19], [34], [95] show that it suffers from overfitting due to using tests as an oracle for repairs, which may not generalize to unseen tests.

BeAFIX [34] addresses the overfitting issue by using assertions as the basis for repairing Alloy specifications. Assertions are more natural in Alloy development and represent a larger set of tests than individual test cases. BeAFix also relies on the user to identify faulty statements. To generate repairs, BeAFix exhaustively searches all possible repair candidates up to a certain bound by mutation and uses Alloy counterexamples for variabilization feasibility checking.

ATR [95] is a template-based approach for repairing Alloy specifications that extends the fault localization techniques of FLACK. FLACK is designed to identify potentially faulty Alloy expressions responsible for assertion failures by analyzing counterexamples and PMAXSAT instances to extract relevant information. ATR builds on FLACK's methodology by generating repair candidates using templates and by leveraging the discrepancies between counterexamples and satisfying instances to eliminate unsuitable candidates.

ICEBAR [19] is a tool that uses ARepair as a backend tool for repairing faulty Alloy specifications based on a set of Alloy tests. However, ARepair is more susceptible to producing overfitting repairs due to its use of tests as oracles. ICEBAR mitigates overfitting by using property-based oracles, which represent large families of intended behavior tests. It also uses Alloy counterexamples to prune certain partial fixes that can never lead to repairs.

*B. Large Language Models*

Large Language Models (LLMs) [20] have been pre-trained [22], [71] on large text corpora so that they learn the general knowledge about the target natural language. Then, these models are either fine-tuned [71] with task-specific data or prompted [52] with task-related information to obtain a decent performance on various downstream tasks. Fine-tuning requires a human annotated dataset to train the LLMs in a supervised manner. Since constructing a training dataset with sufficient size is costly, prompting may be a desired alternative for LLMs with billions of parameters. It directly instructs LLMs without any fine-tuning by providing natural language descriptions of the downstream task and optionally a few examples of the task being handled by human, which relieves the cost of collecting human annotated training data in finetuning.

LLMs are built on the transformer architecture [86] and can be classified into auto-regressive language models [20], masked language models [25], and encoder-decoder [48] language models based on their pre-training objectives. Autoregressive language models employ "generative pre-training", predicting the next token in a sequence (e.g., GPT-3 [20], [73]). Masked language models predict masked tokens using the remaining token sequence as context (e.g., BERT [25]), while encoder-decoder language models are trained on sequence-to-sequence tasks, excelling in translation and summarization tasks (e.g., T5, BART [48], [72]). Recently, pre-trained natural language LLMs have been adapted for source code, such as CodeGPT/Codex [21] (based on GPT3 [20]), CodeBERT/GraphCodeBERT [26], [33] (based on BERT [25], [53]), and CodeT5 [90] (based on T5 [72]). Those LLMs have been evaluated and fine-tuned for various software engineering tasks, such as automated code repair and clone detection, demonstrating promising results.

More recently, researchers have found that LLMs trained under a reinforcement learning paradigm tend to align better with human preference [68], [69], [96]. Specifically, ChatGPT [68] is a LLM that was trained in such a paradigm and has received much attention from researchers and practitioners. It is first initialized from a pre-trained model on auto-regressive generation and then fine-tuned using reinforcement learning from human feedback (RLHF) [96]. RLHF first fine-tunes the model with a small dataset of prompts and desired outputs written by human labelers. Next, a reward model is trained on a larger set of prompts by sampling outputs from the fine-tuned model and using human labeler to rank each individual output, which reflects the human preference. Finally, reinforcement learning is applied to calculate the reward of the generated output based on the reward model and adjust the weights of LLM. The trained LLMs (e.g., ChatGPT) have shown better understanding of human intentions and superior performances in various downstream tasks [16], [40], [49], [69], [91] to most state-of-the-art LLM-based approaches.

However, there is still a lack of understanding about ChatGPT's feasibility and performance on automated specification repair. In this work, we make the first step to evaluate ChatGPT's ability to improve the correctness and completeness of Alloy declarative specifications through automatic repairs.

## IV. EXPERIMENTAL DESIGN

*A. Repair Scenarios*

We explore five repair scenarios that involve different prompts given to ChatGPT to guide the repair process. These repair scenarios are intended to cover all practical repair situations similar to those studied in prior APR research. A ChatGPT prompt refers to the initial text or input given to the ChatGPT language model, which is designed to generate responses. It is a starting point for the AI system to generate a coherent and relevant response based on its training and knowledge. The prompt can be a question, statement, or any other form of input that the user provides to initiate a conversation or request information from the ChatGPT language model. In the following we present each of the five repair scenarios, summarized in Table I, along with a concrete example, depicted in in Figure 2.

- **Repair Scenario 1**. The benchmark specifications contained both bug locations and fix comments. Additionally, ChatGPT was tasked with fixing the bugs without being instructed to pass the assertions.
- **Repair Scenario 2.** The benchmark specifications only included bug locations, without any fix comments. Additionally, ChatGPT was tasked with fixing the bugs without being instructed to pass the assertions.
- **Repair Scenario 3.** The benchmark specifications had no bug locations or fix comments. Additionally, ChatGPT was tasked with passing the assertions to fix the bugs.
- **Repair Scenario 4.** The benchmark specifications had no bug locations or fix comments. Additionally, ChatGPT was tasked with fixing the bugs without being instructed to pass the assertions.
- **Repair Scenario 5.** The benchmark specifications only included bug locations, without any fix comments. Additionally, ChatGPT was tasked with passing the assertions to fix the bugs.

```
Repair Scenario 1: "Consider the following Alloy speci-
fication and its inline comments. There is a bug in this
specification. Find the bug and fix it."

pred ObjectNoExt() {
  //Object does not extend any class.
  //Fix: replace "c.^ext" with "c.^~ext"
  //or "c.~^ext".
  all c: Class | Object !in c.^ext
}
```

```
Repair Scenario 2: "Consider the following Alloy speci-
fication and its inline comments. There is a bug in this
specification. Find the bug and fix it."

pred ObjectNoExt() {
  // Object does not extend any class.
  // Bug:
  all c: Class | Object !in c.^ext
}
```

```
Repair Scenario 3: "Consider the following Alloy speci-
fication and its inline comments. There is a bug in this
specification. Fix the bug so that it satisfies the assertions."

pred ObjectNoExt() {
  // Object does not extend any class.
  all c: Class | Object !in c.^ext
}
```

```
Repair Scenario 4: "Consider the following Alloy speci-
fication and its inline comments. There is a bug in this
specification. Find the bug and fix it."

pred ObjectNoExt() {
  // Object does not extend any class.
  all c: Class | Object !in c.^ext
}
```

```
Repair Scenario 5: "Consider the following Alloy speci-
fication and its inline comments. Fix the bug so that it satisfies the assertions."

pred ObjectNoExt() {
  // Object does not extend any class.
  // Bug:
  all c: Class | Object !in c.^ext
}
```

Fig. 2: Examples of prompts for various repair scenarios. Figures partially represent excerpts from the faulty specification provided to ChatGPT.

| Repair Scenario | Bug location | Fix comment | Passing assertion requirement |
|---|---|---|---|
| Scenario-1 | Yes | Yes | No |
| Scenario-2 | Yes | No | No |
| Scenario-3 | No | No | Yes |
| Scenario-4 | No | No | No |
| Scenario-5 | Yes | No | Yes |

TABLE I: Repair Scenarios Summary

*B. Baselines*

In this study, we compare a ChatGPT-based Alloy repair technique with four state-of-the-art automated repair techniques for faulty Alloy specifications. These techniques include (1) ARepair [88], the first APR technique for Alloy, (2) BeAFIX [34], a bounded exhaustive repair method, (3) ATR [95], which uses templates for repair, and (4) ICEBAR [19], a feedback-driven iterative repair approach. These techniques have been proposed recently in the literature and serve as baselines for comparison with a ChatGPT-based repair approach. Further details about these baselines can be found in Section III.

*C. Subject Systems*

In our evaluation, we employ two distinct benchmark suites: ARepair [88] and Alloy4Fun [55]. These benchmark suites have been extensively studied and were developed by independent research groups, allowing for a fair comparison of various techniques.

The Alloy4Fun benchmark [55] includes a collection of 1936 hand-written faulty specifications that were gathered from student submissions for six different Alloy problems. These problems include the modeling of a labeled transition system (*lts*), an automated production line (*production*), class registrations (*classroom*), a work and source distribution problem (*cv*), various graph properties such as acyclic and completeness (*graphs*), and a file system trash can (*trash*).

The ARepair benchmark [88] is composed of 38 faulty specifications that were obtained from a total of 12 Alloy problems. Six of these problems, named *addr* (address book modeling), *cd* (object and class hierarchy modeling), *ctree* (undirected tree modeling), *farmer* (chicken crossing problem modeling), *bempl* and *other* (both modeling the security lab access problem), and *grade* (gradebook specification modeling), were obtained from the Alloy Analyzer. The remaining faulty specifications were obtained from graduate student homework and are named *arr* (sorted arrays modeling), *blancedBST* (balanced binary search tree modeling), *dll* (doubly linked list modeling), *fsm* (finite state machine modeling), and *student* (sorted linked list modeling).

The benchmark datasets include specifications ranging from tens to hundreds of lines, with real bugs written by humans. The defects present in the benchmarks cover a diverse range, from straightforward ones that can be fixed by modifying a single operator to complex ones that demand the synthesis of novel expressions and substitution of the entire predicate body. Additionally, the benchmarks come with the correct versions of the specifications, which serve as ground truths for verifying the accuracy of the obtained results.

*D. Implementation*

We implement an apparatus to conduct our empirical study in Java and Python on top of Alloy Analyzer APIs and by accessing the ChatGPT API endpoint, which allowed us to directly collect and save responses. We utilized the gpt-3.5turbo model from the ChatGPT family, which is a stable and reliable version of the GPT engine. Our implementation allows for the direct submission of each faulty Alloy specification prepared for the scenarios detailed earlier via the API and stores the response in a text file.

For the evaluation and verification process of each repaired Alloy specification, we used Alloy Analyzer version 4.2 APIs. We collected responses from ChatGPT five times for each faulty specification in each of the two benchmark suites, based on the aforementioned five scenarios. This involved approximately 10,000 requests sent to the ChatGPT server. To evaluate the generated patches, we used a system with an 8core CPU, 8-core GPU, and 16-core neural engine, all based on the ARM architecture, with 8 GB of RAM and running macOS Ventura with Oracle Java SE Development Kit 8u202 (64-bit version).

For language models such as ChatGPT, the "temperature" is a hyperparameter that controls the randomness of the model's output during text generation. It determines how creative or deterministic the model's responses will be. Choosing an appropriate temperature value depends on the specific use case and the desired output style. So it is essential to systematically explore different values of temperature before deciding. We conducted tests with various temperature values for ChatGPT during our experimentation. We explored temperature values from 0.05 to 1, with a 0.05 scale increment, and compared the results with the samples selected from the benchmark dataset. We found that the default temperature value provided the best results when evaluated on the benchmark samples. This outcome affirmed the effectiveness of the default setting for ChatGPT in our specific task and led us to use it for our experiments.

*E. Methods*

This section overviews the methods we utilized to address the four research questions.

*1) RQ1–Comparing with the state-of-the-art techniques for Alloy specification repair:* To investigate RQ1, we conducted a comparative analysis between ChatGPT-based Alloy repair and four state-of-the-art Alloy repair tools: ATR [95], ARepair [87], BeAFix [34], and ICEBAR [19]. We used the same benchmark specifications, namely the ARepair and Alloy4Fun benchmark suites, that were employed by these tools to ensure a fair evaluation. In both benchmarks, we utilized the specifications' corresponding assertions as oracles and automatically generated test suites using AUnit [84] for ARepair [88]. Recall that ARepair requires tests as oracles during the repair process; we thus followed the procedure suggested in [88], as test cases are not commonly available alongside Alloy specifications. To encompass repair situations commonly explored in prior automated program repair (APR) research, we created and executed five distinct repair scenarios utilizing ChatGPT. These scenarios involve diverse prompts given to ChatGPT, which guide the repair process and are detailed in Section IV-A.

*2) RQ2–Consistency of the repairs:* To assess ChatGPT's reliability as a repair technique, we conducted a consistency test, as LLM-based techniques may not generate the same repairs given the same prompt and faulty specification. The test involved providing ChatGPT with the exact same repair prompt ten different times for ten different faulty specifications, including a bug location indicator. We used new chat sessions each time and measured the similarity of the generated repairs using BLEU, ROUGE-L, METEOR, and BERTbased semantic similarity metrics. We also verified the validity of the generated repairs across ten prompts by checking them using the Alloy formal analyzer, ensuring consistency.

*3) RQ3–Category of Mistakes:* For answering RQ3, we did a manual inspection of the mistakes made by ChatGPT. In total, we analyzed 6,559 mistakes. We followed an open coding protocol [29] to categorize the mistakes. Specifically, two authors of this study independently checked all mistakes and categorized them. During the analysis, each emerging category was compared with existing ones to determine if it is a new category through multiple comparison sessions. Finally, the two authors exchanged ideas for the categorization and reached a consensus through negotiated agreement [27].

*4) RQ4–Impacts of Prompts:* To investigate the impact of prompts on the Alloy repairs produced by ChatGPT, we designed a study that varies three main elements of prompts for Alloy repair: (1) Bug Location Indicator: A binary factor that determines whether the prompt contains a bug location indicator or not. (2) Fix Comment: A binary factor that determines whether the prompt contains a fix comment or not. (3) Assertion Instruction: A binary factor that determines whether the prompt instructs ChatGPT to pass assertions or not. We also followed the best practices outlined in [2]. We designed a set of repair prompts based on the above factors (cf. Section IV-A and Fig. 2). For each prompt, we ran the ChatGPT repair process and collected the resulting repairs. We evaluated the quality of the repair using the same metrics as in our prior experiments. We also manually inspected the repairs to identify any major mistakes or hallucinations.

## V. EXPERIMENTAL RESULTS

**Results for RQ1: State-of-the-art Comparison.**

In our experiments, we compared the ChatGPT-based Alloy repair with four state-of-the-art Alloy repair tools: ATR [95], ARepair [87], BeAFix [34] and ICEBAR [19]. To ensure a fair evaluation, we employed the same benchmark specifications as those used by these tools. The results of our findings are displayed in Table II. The columns labeled Models and Total #specs indicate the model groups and the total number of buggy specifications in each group, respectively. Each subsequent column in the table represents the total number of correctly repaired specifications by each respective tool. It is worth mentioning that each tool has a different development strategy and addresses specification attributes differently. Overall, ChatGPT was able to repair 26% (517/1974) of the total specifications in Repair Scenario 2. This represents the best result achieved by ChatGPT in this scenario (see Fig 2). In the individual benchmarks, ChatGPT repaired 26% (497/1936) of the Alloy4Fun benchmark in Repair Scenario 2 and 76% (29/38) of the ARepair benchmark in Repair Scenario 1. In the overall comparison of repair scenarios 1, 2, 3, and 5, ChatGPT outperformed ARepair, which only had a 9% repair rate among all the specifications. Although ChatGPT falls behind the other repair tools in the Alloy4Fun benchmark, except for ARepair, it achieved the best performance on the ARepair benchmark with 29 repairs in Repair Scenario 1.

TABLE II: Comparison to the state-of-the-art Alloy repair techniques, i.e., ARepair [88], ICEBAR [19], BeAFIX [34], and ATR [95], on the ARepair and Alloy4Fun benchmarks.

| | Model | Total #specs | Arepair #repairs | ICEBAR #repairs | BeAFix #repairs | ATR #repairs | Scenerio-1 #repairs | Scenerio-2 #repairs | Scenerio-3 #repairs | Scenerio-4 #repairs | Sceneirio-5 #repairs |
|---|---|---|---|---|---|---|---|---|---|---|---|
| Alloy4Fun | classroom | 999 | 88 | 424 | 387 | 688 | 139 | 231 | 94 | 88 | 162 |
| | cv | 138 | 2 | 86 | 82 | 38 | 58 | 50 | 43 | 4 | 53 |
| | graphs | 283 | 19 | 237 | 232 | 260 | 78 | 109 | 90 | 20 | 75 |
| | lts | 249 | 1 | 73 | 41 | 70 | 91 | 70 | 49 | 21 | 53 |
| | production | 61 | 27 | 36 | 56 | 43 | 28 | 32 | 24 | 12 | 26 |
| | trash | 206 | 48 | 195 | 183 | 187 | 7 | 5 | 3 | 2 | 5 |
| | **Summary** | **1936** | **185** | **1051** | **981** | **1286** | **401** | **497** | **303** | **147** | **374** |
| ARepair | addr | 1 | 1 | 1 | 1 | 1 | 1 | 0 | 0 | 0 | 1 |
| | arr | 2 | 2 | 2 | 2 | 1 | 1 | 1 | 1 | 0 | 1 |
| | balancedBST | 3 | 1 | 2 | 1 | 1 | 3 | 2 | 2 | 0 | 0 |
| | bempl | 1 | 0 | 1 | 0 | 1 | 0 | 1 | 1 | 1 | 1 |
| | cd | 2 | 0 | 2 | 2 | 2 | 1 | 1 | 2 | 0 | 2 |
| | ctree | 1 | 1 | 0 | 0 | 0 | 0 | 0 | 1 | 0 | 1 |
| | dll | 4 | 0 | 3 | 3 | 2 | 4 | 4 | 3 | 0 | 1 |
| | farmer | 1 | 0 | 0 | 0 | 0 | 1 | 0 | 0 | 0 | 0 |
| | fsm | 2 | 2 | 2 | 1 | 2 | 2 | 1 | 0 | 0 | 0 |
| | grade | 1 | 0 | 1 | 0 | 1 | 1 | 0 | 1 | 0 | 0 |
| | other | 1 | 0 | 0 | 1 | 1 | 1 | 0 | 0 | 1 | 0 |
| | student | 19 | 2 | 7 | 13 | 10 | 14 | 10 | 15 | 2 | 4 |
| | **Summary** | **38** | **9** | **21** | **24** | **22** | **29** | **20** | **26** | **4** | **11** |
| | **Total** | **1974** | **194** | **1072** | **1005** | **1308** | **430** | **517** | **329** | **151** | **385** |

TABLE III: Concurrent Representation of Similarity Metrics for ChatGPT's Repairs in Consistency Test. ChatGPT was prompted with ten repair prompts, each repeated ten times, and the similarity of its repairs for each prompt was measured.

| | **Instructed to Pass Assertion** | | | | **Not instructed to Pass Assertion** | | | |
|---|---|---|---|---|---|---|---|---|
| **PROMPT** | BLEU | ROUGE-L | METEOR | BERTSim | BLEU | ROUGE-L | METEOR | BERTSim |
| 1 | 0.642 | 0.281 | 0 | 0.641 | 0.139 | 0.245 | 0 | 0.585 |
| 2 | 0.906 | 0.708 | 0.613 | 0.905 | 0 | 0 | 0.135 | 0.525 |
| 3 | 0.481 | 0.062 | 0 | 0.481 | 0 | 0 | 0.646 | 0.537 |
| 4 | 0.982 | 0.857 | 0.646 | 0.982 | 0 | 0 | 0.113 | 0 |
| 5 | 0.859 | 0.562 | 0.539 | 0.858 | 0.034 | 0.058 | 0.135 | 0.593 |
| 6 | 0.718 | 0.125 | 0 | 0.717 | 0 | 0 | 0.646 | 0 |
| 7 | 0.624 | 0.062 | 0.646 | 0.623 | 0 | 0 | 0 | 0 |
| 8 | 0.859 | 0.562 | 0 | 0.858 | 0.121 | 0.201 | 0.060 | 0.647 |
| 9 | 0.859 | 0.562 | 0 | 0.858 | 0 | 0 | 0 | 0.537 |
| 10 | 0.718 | 0.432 | 0 | 0.717 | 0.034 | 0.058 | 0.189 | 0.593 |
| **Average** | **0.765** | **0.378** | **0.244** | **0.764** | **0.033** | **0.0562** | **0.202** | **0.402** |

While ChatGPT did not surpass the existing tools in terms of total repairs, it demonstrated promising results by fixing unique specifications that were not addressed by the previous tools. For example:

- Specification *bempl* could not be fixed by ARepair and BeAFix but was successfully repaired by ChatGPT in 4 instances.
- Specifications *cd* and *dll* could not be fixed by ARepair, but ChatGPT fixed them in 4 instances.
- The specification *farmer* remained unfixed by all four repair tools but was resolved by ChatGPT in Repair Scenario 1.
- Specification *grade* could not be fixed by ARepair and BeAFix, but ChatGPT resolved it in 2 instances.
- Specification *other* could not be fixed by ARepair and ICEBAR, but ChatGPT fixed it in 2 instances.

Based on the results, ChatGPT exhibits inconsistent and varied behavior across the benchmarks. While it does not consistently outperform the existing tools, it demonstrates promise in finding fixes for specific buggy specifications that the other tools fail to address. This indicates that ChatGPT has potential for resolving problematic specifications.

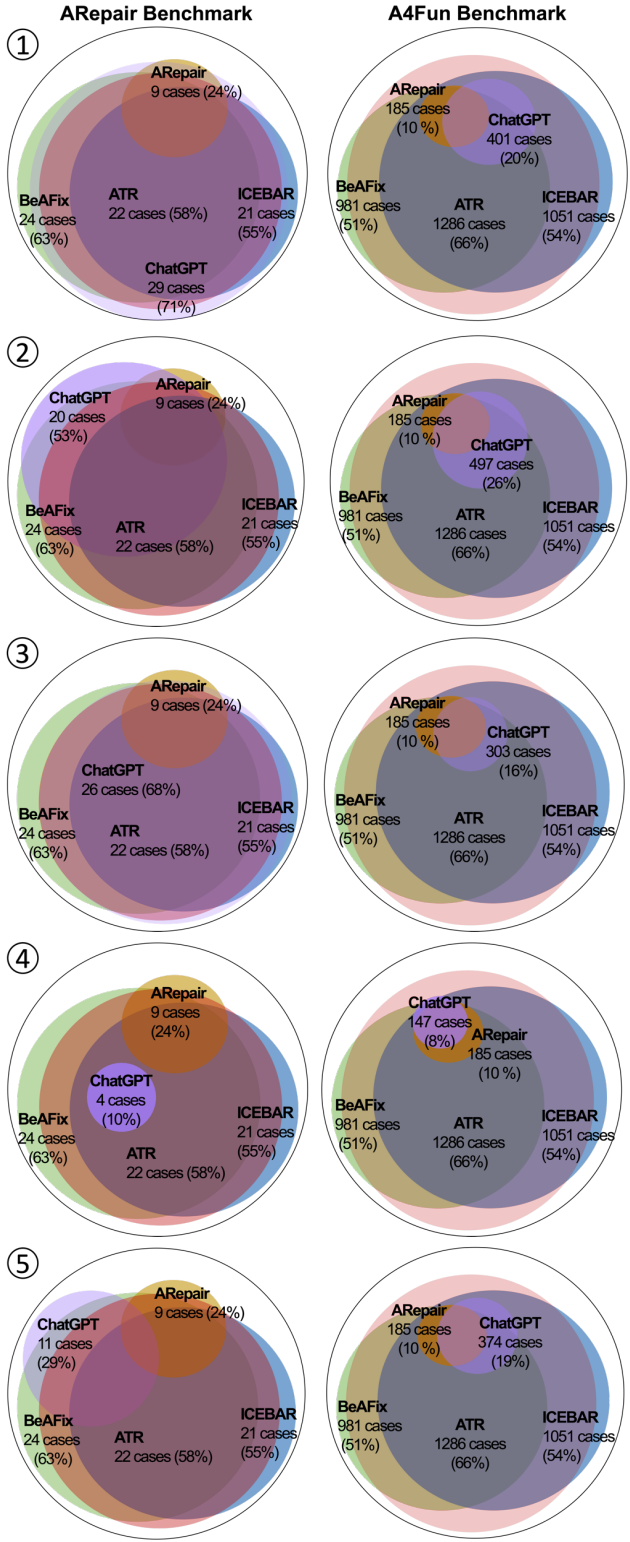

Fig. 3: Venn diagrams depict the comparison of ARepair, ICEBAR, BeAFix, ATR, and ChatGPT in five different scenarios across two benchmark suits. The repair cases shared between these techniques are highlighted in the overlapping regions.

**Observation 1: ChatGPT can fix unique specifications that were not addressed by the previous repair tools.**

### Results for RQ2: Consistency of the repairs.

One potential issue with repair techniques based on LLMs is that, given the exact prompt and faulty specification, they may not generate the same repairs. So we conducted a consistency check to evaluate ChatGPT's reliability as a repair technique by assessing whether it consistently generates the same repairs given the same prompt and faulty specification. Ten different faulty specifications were used, and for each one, we provided the same repair prompt ten times to ChatGPT. Each prompt was given in a new chat session with no context, and we asked ChatGPT to satisfy the assertions for each of these faulty specifications: "Consider the following Alloy specification. There is a bug in this specification. Fix the bug so that it satisfies the assertions." We then conducted ten repetitions of the same prompt on ChatGPT for the same ten faulty specifications. Each prompt was again presented in a new chat session with no context. However, during these replications, ChatGPT was not explicitly requested to satisfy the assertions.

We measured the similarity of ChatGPT's generated repairs using various metrics such as BLEU, ROUGE-L, METEOR, and a BERT-based semantic similarity metric. The results, presented in Table III, show the average similarity among the ten generated repairs for each of the ten prompts. For instance, when ChatGPT was instructed to pass the assertion for prompt 1, the ten instances of repairs had an average score of 0.642 on BLEU, 0.281 on ROUGE-L, 0 on METEOR, and 0.641 on the BERT-based semantic similarity metric. The BERT-based semantic similarity metric is particularly meaningful as it evaluates repairs in a semantic space by converting them into BERT encodings. However, the obtained results indicate that ChatGPT's responses to the same prompts are not highly consistent. We used Welch's t-test [76] to check if the scores for "Instructed to pass assertion" and "Not instructed to pass assertion" are statistically significantly different or not for the same prompt. We decided to use Welch's t-test because it does not assume equal variance between groups but only assumes normality of the data. Considering the multiple tests conducted, we applied the Bonferroni correction [35] to account for multiple hypothesis corrections, resulting in an adjusted p-value of 0.005. The analysis results revealed significant differences between all pairs of scores for the same prompt (Welch's t-test, p-val<6.660e-05).

**Observation 2: ChatGPT shows inconsistency by generating different repairs for the same input on different executions.**

### Results for RQ3: Primary Category of Mistakes.

During our assessment of the ChatGPT's unsuccessful repairs, we discovered that the responses across different repair scenarios contained a number of errors in the Alloy specifications. Upon further investigation, we identified several interesting categories of these mistakes. Table IV displays the categories of mistakes found in ChatGPT repairs across the five repair scenarios, along with their respective quantities and proportions within each category.

- *Incorrect use of operator*: We observed a few errors where incorrect operators were used. For instance, the **"&"** operator was used in certain cases where both sides of an expression were not compatible. This mismatch resulted in a type error. Similar issues led to other errors as well.

TABLE IV: Quantity and proportion of mistakes from ChatGPT repairs

| Type of Errors | Quantity and proportion of mistakes | | | | |
|---|---|---|---|---|---|
| | Scenario-1 | Scenario-2 | Scenario-3 | Scenario-4 | Scenario-5 |
| Wrong use of operator | 336/1974 (17%) | 197/1974 (10%) | 278/1974 (14%) | 120/1974 (6.07%) | 158/1974 (8%) |
| Wrong type | 450/1974 (23%) | 246/1974 (12%) | 424/1974 (21%) | 1125/1974 (57%) | 637/1974 (32%) |
| Misusing higher-order logic | 0 (0%) | 0 (0%) | 0 (0%) | 1 (0.05%) | 0 (0%) |
| Mismatch relational arity | 321/1974 (16%) | 349/1974 (18%) | 335/1974 (17%) | 225/1974 (11%) | 332/1974 (17%) |
| Hallucination | 180/1974 (9%) | 122/1974 (6%) | 162/1974 (8%) | 186/1974 (9%) | 370/1974 (19%) |

- *Incorrect type*: This error occurred due to the presence of unexpected tokens such as `" "`, `"{"`, and `"|"`. We noticed these errors when the structure or arrangement of the code did not conform to the grammar and rules of the Alloy specifications, resulting in syntax errors.
- *Misuse of higher-order logic*: We encountered this type of logical error, which happened due to the absence of higher-order quantification. This occurred in a single case. This specific error suggests that higher-order quantification could not be skolemized, meaning that the process of replacing quantified variables with skolem constants or functions was unsuccessful. Consequently, the analysis could not proceed, resulting in an error.
- *Mismatch in relational arity*: This error occurred when a relational join was not legal or valid due to incompatible types on the left-hand side and right-hand side of the expression. For example, consider the following expression:

```
left-hand side is u.(this/User <: profile)
(type = {this/Work})
right-hand side is this/Work
(type = {this/Work})}
```

In this expression, the specific relational join provided by ChatGPT is not valid because the left-hand side and the right-hand side have incompatible types. The left hand side involves the subset operator `<:` applied to this/User and profile, while the right-hand side is simply this/Work. We observed such issues in a large number of instances provided by ChatGPT, resulting in Type Errors.

• *Hallucination*: This error occurred when ChatGPT referenced a variable in a predicate or fact that had not been previously defined or declared within the scope of the current code.

Based on our thorough investigation, we were able to categorize these mistakes across all repair scenarios. Among the error types, the **Incorrect type** had the highest impact, affecting (57%) of the specifications in Repair Scenario 4 and (32%) in Repair Scenario 5 across the two benchmarks. Although the **Missing higher-order logic** error only affected one specification, it provided significant insight into the types of mistakes that can occur with ChatGPT. The error type **Hallucination** indicates that ChatGPT's attempts to repair the program reflect its large-scale training capabilities, but also highlight the imperfections in its assumptions. The error types **Incorrect use of operator and Incorrect type** indicate that ChatGPT's logic representation for repairing a specification might be initially correct, but it becomes inaccurate due to the use of unnecessary or irrelevant operators. The errors made by ChatGPT demonstrate that the repairs provided may lack reliability and consistency. The categorization of errors based on each repair scenario can be found in Table IV.

> **Observation 3: ChatGPT makes different types of mistakes with Incorrect type being the most frequent.**

**Results for RQ4: Impact of the prompts.**

The Venn diagrams presented in Figure 3 illustrate the outcomes obtained by ChatGPT in five distinct scenarios across two benchmark suites. This depiction showcases the interrelation and commonality of solutions among various tools. As discussed in RQ1, ChatGPT demonstrates the ability to repair certain unique faulty specifications in the ARepair benchmark suite that were not addressed by any of the other repair techniques. With the exception of *Repair Scenario 4*, ChatGPT's repairs were not entirely encompassed by all the other repair techniques in the ARepair benchmark suite. In this specific scenario, the repairs generated by ChatGPT aligned completely with those produced by ICEBAR, BeAFix, and ATR. However, the results appear inconsistent when attempting to *satisfy the assertion* in *Repair Scenarios 3* and *5*. In *Repair Scenario 3*, the results overlap with the other tools, while in *Repair Scenario 5*, the number of correct fixes is comparatively lower. Nonetheless, there are some distinctive fixes for specific specifications like *bempl*, *cd*, and *dll*. When considering the Alloy4Fun benchmarks, the results across all repair scenarios were found to be a subset of ICEBAR, BeAFix, and ATR. The Venn diagrams provide a clear illustration of how the impact of different prompts has a diverse effect on the segmentation of the data and influences ChatGPT's responses to each scenario. Specifically, the obtained results reveal a significant impact of the prompt, with *Repair Scenario 4*, which has the least amount of guidance provided, performing the worst compared to all other *Repair Scenarios*.

During the analysis of the impact of "Bug Location" information as part of the prompt, the bug location was provided in the specification as a comment, but only in specific repair scenarios (cf. Table I) for both benchmark suites. Notably, in the ARepair benchmark, there was clear guidance on how to fix the bug in one scenario (i.e., Scenario 1), as shown in Figure 2 (Repair Scenario 1), for instance, in the form of "replace c.ˆext with c.ˆ~ext". However, for the Alloy4Fun benchmark, there was no explicit instruction on how to conduct the fix. Consequently, the performance of ChatGPT for the Alloy4Fun benchmark consistently lagged behind that of the ARepair benchmark, indicating the impact of "Bug Location."

When analyzing the impact of assertion information, we observed no significant impact of requiring to pass assertions in the prompt. We posit that this may be attributed to ChatGPT's proficiency in understanding natural language text but limited capability in verifying logical assertions.

> **Observation 4**: Specific instruction on how to fix the bug as part of the prompt has the most significant impact on ChatGPT's performance.

## VI. DISCUSSION AND IMPLICATIONS

This section presents implications for tool builders, researchers, and software practitioners when considering the use of OpenAI's ChatGPT for automatic repair of declarative specifications.

**Implications for software practitioners.** ChatGPT falls short when compared to existing techniques in terms of overall effectiveness. Therefore, Alloy users should not solely rely on ChatGPT for all repair tasks. Rather, users should consider ChatGPT as a supplementary tool for specification repair, particularly when other techniques have exhausted their capabilities or failed to address certain types of bugs, as indicated in Observation 1. In such cases, ChatGPT may provide alternative solutions that can be considered alongside existing repairs.

Due to identified errors in ChatGPT's generated repairs, users should exercise caution and critically validate the suggested repairs. Specifically, users should be aware of potential issues like improper operator usage, type errors, higher-order logic misuse, relational arity mismatches, and hallucinations when relying on ChatGPT for specification repairs. The most frequent errors observed are related to the wrong use of operators and types. This occurrence can be attributed, in part, to Alloy's alternative syntax for various logic operators (e.g., 'and' for conjunction) and the syntactic similarities between certain logical and relational operators (e.g., 'not' for negation vs. 'no' for emptiness check, '&&' for conjunction vs. '&' for intersection). Manually reviewing and verifying the repairs is essential to ensure their correctness and reliability.

**Implications for researchers.** Our study highlights that ChatGPT can fix unique specifications that were not addressed by state-of-the-art repair tools. This suggests that ChatGPT might be particularly useful for niche or specific buggy specifications that might be challenging for conventional repair approaches. Further exploration and understanding of the types of buggy specifications ChatGPT is most effective at repairing can help in optimizing its use.

Our study identifies instances of hallucinations in generated repairs. Future research should focus on minimizing such hallucinations through improved training, fine-tuning, and validation processes. Inconsistency in the repairs generated by ChatGPT is another area that needs attention. Ensuring more consistent and reliable repairs can be achieved by refining the model's behavior in specific scenarios and enhancing its ability to produce consistent repairs to similar buggy specifications.

**Implications for tool builders.** Despite falling short in comparison to existing repair techniques for Alloy declarative specifications, ChatGPT has demonstrated its ability to address certain buggy specifications that existing methods could not handle. This indicates that it can be used as a complementary tool alongside other repair techniques to improve overall effectiveness. For example, a hybrid approach combining existing tools and ChatGPT is worth exploring.

Given the identified errors and inconsistencies, we posit that incorporating a human-in-the-loop validation step in the repair process with ChatGPT may be able to enhance the reliability of the repaired specifications. Indeed, while it may be useful for brainstorming alternative solutions to challenging specification issues, human validation backed by the automated Alloy Analyzer can help verify the correctness of repairs, prevent hallucinations from being accepted, and provide additional insights for further improvement.

## VII. THREATS TO VALIDITY

We have taken all reasonable steps to mitigate potential threats that could hamper the validity of this study, it is still possible that our mitigation strategies might not have been effective.

ChatGPT may show better performance when being used in a conversational way. That is, humans provide necessary feedback to the response ChatGPT gives so that the model can give better results in the following rounds. However, as the first work to investigate the performance and the feasibility of ChatGPT on repairing specification bugs, we believe our analysis sheds light on how good ChatGPT fixes specification bugs in Alloy.

It is possible that our manual labeling process for identifying the categories could have introduced unintentional bias. To address this, two authors inspected and labeled independently. A Cohen kappa of 0.95 indicates a high reliability of our labeling.

Our conclusions may not be generalizable to Alloy problems that are not included in the two benchmarks we experimented in this study.

## VIII. CONCLUSION

To aid programmers in rectifying mistakes in specifications and promote the adoption of formal specifications in day-today software engineering, various automatic repair approaches have recently been proposed. Despite notable progress, these approaches exhibit certain limitations. Recently introduced, ChatGPT, a large language model, can generate fixes for bugs in specifications based on provided prompts and related Alloy specifications.

Our findings reveal that while ChatGPT is not able to perform on par with the existing techniques, it demonstrates surprising capability in repairing bugs that other techniques fail to address. However, our results also identify some mistakes made by ChatGPT, including incorrect operator usage, incorrect types, misuse of higher-order logic, and relational arity mismatches. Additionally, we observed instances of hallucination in the generated repairs. The findings of this study offer valuable insights for tool builders, researchers, and software practitioners who are contemplating the utilization of ChatGPT for automatic repair of declarative specifications.

The data collected and statistical analyses conducted in this study have been made publicly available on the project website [1] to ensure the verifiability and reproducibility of our findings.